\documentclass[aps,prb,showkeys,preprint,showpacs,12pt]{revtex4-1}
\usepackage{graphics,graphicx}
\usepackage{color}
\usepackage{amsbsy,amssymb}
\usepackage{amsmath}
\usepackage{epsfig}
\usepackage{color}
\usepackage{textcomp}
\begin{document}
\title{Magnetodielectric effect of Graphene-PVA Nanocomposites}

\author{Sreemanta Mitra$^{1,2}$}
\email[]{sreemanta85@gmail.com}
\author{Oindrila Mondal$^{3}$}
\author{Dhriti ranjan Saha$^{1}$}
\author{Anindya Datta $^{4}$}
\author{Sourish Banerjee$^{2}$}
\author{Dipankar Chakravorty$^{1,\dag}$}
\email[]{mlsdc@iacs.res.in}
\affiliation{
$^{1}$
 MLS Prof.of Physics' Unit,Indian Association for the Cultivation of Science, Kolkata-700032, India.\\ }
\affiliation{
$^{2}$
Department of Physics, University of Calcutta, Kolkata-700009, India.\\}
\affiliation{
$^{3}$ Department of Physics, M.U.C. Woman's College, Burdwan, India.\\}

\affiliation{
$^{4}$
University School of Basic and Applied Science (USBAS),Guru Govind Singh Indraprastha University,New Delhi, India\\}
\begin{abstract}
Graphene-Polyvinyl alcohol (PVA) nanocomposite films with thickness $120 \mu m$ were synthesized by solidification of PVA in a solution with dispersed graphene nanosheets.
Electrical conductivity data were explained as arising due to hopping of carriers between localized states formed at the graphene-PVA interface.
Dielectric permittivity data as a function of frequency indicated
the occurrence of Debye-type relaxation mechanism. The nanocomposites showed a magnetodielectric effect with the dielectric constant changing by $1.8{\%}$ as the magnetic field 
was increased to 1 Tesla. The effect was explained as arising due to Maxwell-Wagner polarization as applied to an inhomogeneous two-dimensional,two-component composite model. 
This type of nanocomposite may be suitable for applications involving nanogenerators.    
\end{abstract}

\maketitle
\section{Introduction}\label{sec:1}
  Graphene has been at the forefront of nanomaterials research in recent times \cite{geim,novo}. Apart from devising simpler and more efficient ways of synthesizing graphene
sheets \cite{stankovich,lotya,cnrjpc,cnrjmc,cnrwi,reinanl},theoretical work has progressed at a fast pace to explain the intriguing electronic properties of graphene 
\cite{castro,sds,allor,shytov,young,stander,sui}. The elucidation of quantum transport mechanism in graphene will have a strong bearing on the design of graphene based 
nanoscale devices. A number of investigations have been reported regarding synthesis of composites incorporating graphene sheets. Transparent and electrically conducting
graphene-silica composites have been prepared and characterized \cite{watcharotonenl}. Polystyrene-graphene composites formed via complete exfoliation of graphite and 
dispersion of resultant graphene sheets in the polymer have shown a very low percolation threshold for room temperature conductivity \cite{stankovichnat}. Graphene-metal
particle nanocomposites have been synthesized by first adsorbing metal particles on graphene oxide sheets which are then catalytically reduced to graphene \cite{xujpcc}. 
An interesting development is the method of chemical storage of hydrogen in a few-layer graphene structure \cite{cnrpnas}. Graphene-Polyaniline nanocomposites have been
used for hydrogen gas sensing \cite{almashatjpcc}.
\par
Our objective was to investigate the electrical properties of a composite consisting of graphene sheets dispersed in a polymer matrix. Apart from the conduction mechanism
we wanted to explore the possibility of inducing magnetodielectric effect in these materials because of inhomogeneous conductor microstructure within them. It was indeed possible
to obtain such behaviour. The details are reported in this paper.
\section{Experimental}\label{sec:2}
The graphene-PVA nanocomposite was synthesized by a simple chemical polymerization method. Primarily, graphene nanosheets were prepared by the chemical reduction and 
exfoliation of graphite oxide. The graphite oxide (GO) was prepared from high purity graphite flakes (as obtained from LOBACHEMIE) using modified 
Hummers' method \cite{hummersjacs,zhoujpcc}.
In brief, 2gm of powdered flake graphite was added to a mixture of 50ml concentrated sulfuric acid ($98{\%}$,as obtained from E-Merck) 2gm $NaNO_{3}$ and 
6gm $KMnO_{4}$.The ingredients were mixed in a beaker that had been cooled to 273K in an ice-bath. After removing the ice-bath the temperature of the suspension was
 brought to room temperature under constant stirring. After 3 hrs,300ml distilled water and 5ml hydrogen peroxide($H_{2}O_{2}$) were added to the thick mixture under stirring.
 Finally, the suspension was filtered,resulting in a yelloish brown cake. This was washed with 1:10 HCl solution inorder to remove unwanted metal ions \cite{kovtycm}.
 The washed product was then collected and dried in an air oven at 333K. Graphene (or more suitably reduced graphene oxide) was prepared by usual chemical reduction
 technique \cite{dannn}. 0.01gm GO powder was uniformly dispersed in 10 ml water and 6 ml hydrazine hydrate was added. The pH of the mixture was kept at ${\sim 10}$ by adding 
ammonia solution. The mixture was stirred thoroughly at room temperature for 3 hrs to remove the oxygen functionality of the graphite oxide.
\par
\begin{figure}
\includegraphics[width= 8.25cm]{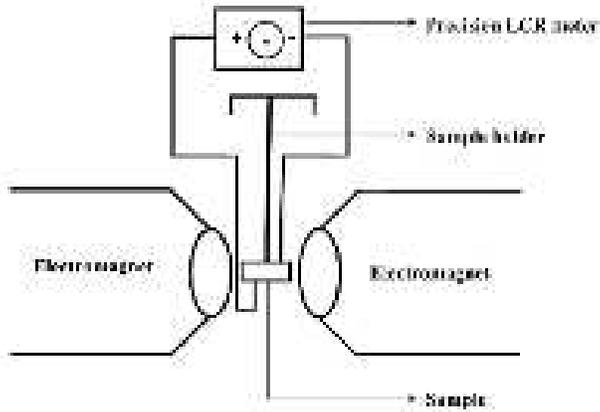}

\caption{Schematic diagram of the experimental set-up for the magneto-dielectric experiment.}

\label{fig.1}
   \end{figure}

0.909 gm of poly-vinyl alcohol (PVA) powder (as obtained from S-d fine-Chem Pvt.Ltd.) was dissolved in 15 ml water and stirred at 333K for 3hrs. to form a
 homogeneous polymer solution. The elevated temperature was needed as PVA does not dissolve in water at room temperature (300K).
It may be noted therefore that the preparation entailed $1{\%}$ GO and $99{\%}$ of PVA (by weight) as precursors. 
3 ml of dispersed graphene  was mixed with 10 ml
PVA solution and stirred for 4 hrs. to form a homogeneous mixture. The mixture was then cast on a teflon petri-dish at ordinary atmosphere. After solidification, 
graphene-PVA nanocomposite film was obtained.
 To characterize the prepared graphene, the sample was mounted on transmission electron microscope (JEOL 2010, operated at 200 kV). Fourier Transform Infrared spectroscopy studies
of both the GO and pristine graphene samples were carried out using FTIR8400S spectrometer. For electrical measurements,silver electrodes (silver paint supplied by M/S Acheson 
Colloiden, B.V. The Netherlands) were applied on both the faces of a piece of graphene-PVA nanocomposite film. For dc electrical resistivity measurement Keithley 617 
electro meter was used,whereas, for dielectric and magnetodielectric coupling measurements, an Agilent E4890A precision LCR meter was used. In order to determine the 
change of dielectric constant with applied magnetic field (magnetodielectric coupling measurement) the orientation of magnetic field was kept perpendicular to the applied 
electric field. In this case magnetic field was generated by a large water cooled electromagnet supplied by M/S Control Systems \& Devices, Mumbai,India. 
The schematic diagram of the experimental set-up has been shown in \ref{fig.1}.


\section{Results and Discussion}\label{sec:3}
\begin{figure}
\includegraphics[width= 8.25cm]{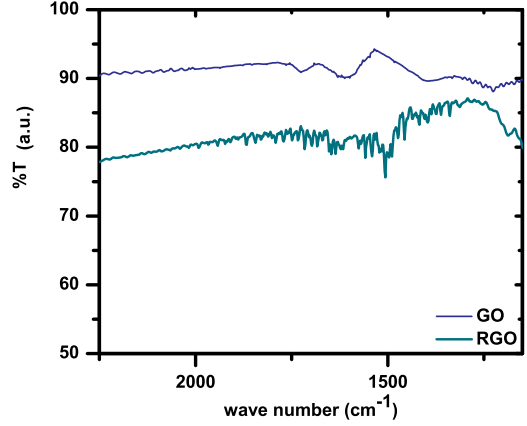}
\caption{(Color online) FTIR spectrograph of both GO and chemically derived graphene.}
\label{fig.2}
   \end{figure}

 The FTIR (Fourier Transform Infrared Spectroscopy) spectra of both the graphite oxide (GO) and the graphene sample has been shown in \ref{fig.2}. 
The spectrum for GO shows transmittance dips at 1390.65 $cm{^{-1}}$,1634.23 $cm{^{-1}}$ 
and 1725.14 $cm{^{-1}}$, which correspond to deformation of O-H bond in water, stretching mode of carbon carbon double bond (C=C) and stretching mode of carbon oxygen double bond 
(C=O) respectively.On the other hand in the spectrum for chemically derived graphene the absence of any dip around 1725 $cm{^{-1}}$ establishes that the derived graphene 
samples are free from the oxygen functionality present in GO.

Figure 3(a) shows the low magnification transmission electron micrograph of the graphene sheet showing the morphology. The folded edge signifies that the sheet is
 very thin as expected, probably 4-5 layers of graphene. The selected area electron diffraction pattern of the graphene sheet is shown in the inset 
of figure 3(a). The high resolution transmission electron micrograph of graphene sheet showing the lattice planes, has been shown in figure 3(b). It should be evident 
that the lattice pattern is hexagonal in nature ,which is further substantiated by the fast fourier transform (FFT) of the image, shown  as inset in figure 3(b).

 \begin{figure}
\includegraphics[width= 8.25cm]{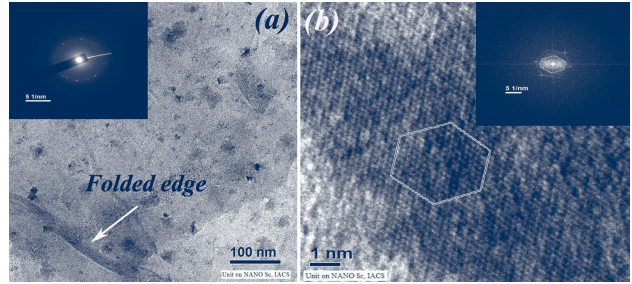}
\caption{(a)A folded edge graphene sheet, (inset)Selected area electron diffraction pattern of (a),(b)High resolution electron micrograph,(inset)FFT image of (b).}
\label{fig.3}
   \end{figure}

\begin{figure}
\includegraphics[width= 8.25cm]{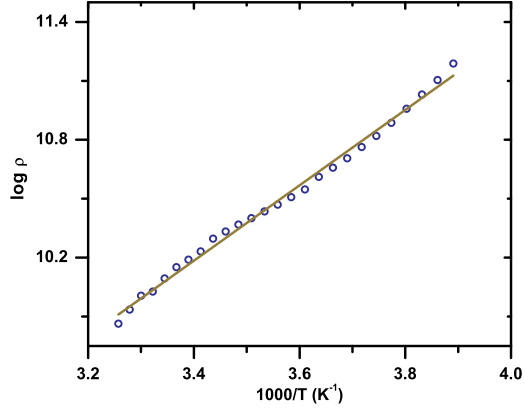}
\caption{(Color online) Variation of logarithm of resistivity with T${^{-1}}$ for the nanocomposite.}
\label{fig.4}
   \end{figure}

The variation of logarithm of dc resistivity as a function of T${^{-1}}$ for the graphene-PVA nanocomposite film has been shown in \ref{fig.4}. 
The silver electroded film has a thickness and area of cross-section ($\sim 0.015 cm $ and $ 0.12 cm{^{2}}$ ) respectively.

From the slope of this Arrhenius plot we calculate a value of activation energy using

\begin{equation}
\textcolor{blue}{\rho=\rho_{o} \exp{(\phi/k_{B}T)}}
  \label{eq.1}
\end{equation}

where, ${\phi}$ is the activation energy which comes out to be  0.38 eV.

The points shown in figure are the experimental points and the solid line represents the least square fitted curve.

\begin{figure}
   \includegraphics[width= 8.25cm]{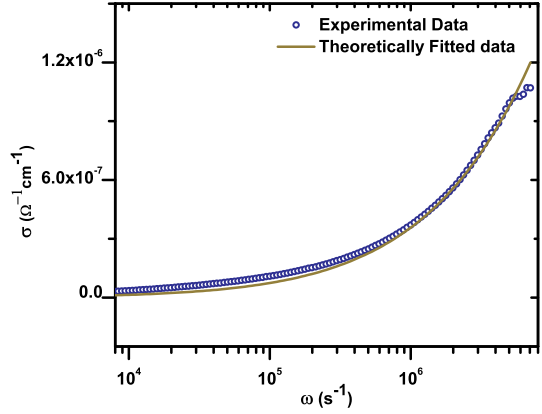}
\caption{(Color online)Variation of the ac conductivity($\sigma(\omega)$) of the nanocomposite as a function of the angular frequency($\omega$) at 304K.}
\label{fig.5}
\end{figure}

 We argue that the edges of the graphene sheets form localised states in the amorphous PVA medium and the conduction mechanism is 
governed by hopping of the carriers near the Fermi level between these localised states \cite{mott}.The variation of the ac conductivity of the nanocomposite as a
function of the angular frequency at 304K is shown in \ref{fig.5}. We have used the formula derived by Davis and Mott, \cite{mott}
  
 \begin{equation}
     \textcolor{blue}{\sigma(\omega)=\frac{{\pi}}{3}{[N(E{_F})]^{2} }{k_{B}}Te^{2}{\alpha}^{-5}{\omega}{[ln(\frac{1}{{\omega}{\tau{_o}}})]^{4}}}
         \label{eq.2}    
    \end{equation}

where $\sigma(\omega)$ is the conductivity for the angular frequency $\omega$,  N$(E{_F})$ is the density of states in the vicinity of Fermi level,
k${_B}$ is Boltzmann constant, T is temperature, e is the electronic charge, $\alpha$ is the localization constant and $\tau{_o}$ is the inverse of phonon frequency.
The experimental data in \ref{fig.5}  were least-square fitted to eq.2 using  N$(E{_F})$,$\tau{_o}$,and ${\alpha}^{-1}$ as parameters. The theoretically fitted curve is 
shown in \ref{fig.5}. The values extracted are N$(E{_F})$=1.0\texttimes10$^{23} eV{^{-1}}cc{^{-1}}$,$\tau{_o}$=1.0\texttimes10$^{-11}$ sec,and ${\alpha}^{-1}$=72 A$^{o}$. 
These values are similar to those reported for various chalcogenide glasses\cite{mott}. 


\par
The dielectric permittivity of the nanocomposite film was measured and we have delineated the real and imaginary components of the same. The variation of real 
${(\epsilon{'})}$ and imaginary ${(\epsilon{''})}$ parts of the dielectric permittivity with the frequency, measured at different temperatures is shown in
figures 6 and 7 respectively. It can be seen that the real part ${(\epsilon{'})}$ decreases from a constant high value as the frequency is increased and on the contrary,
the imaginary part ${(\epsilon{''})}$ shows a Debye-like relaxation peak \cite{hippel},which shifts towards the higher frequency side as the temperature is increased. 
The graphene-PVA nanocomposite film shows dielectric dispersion, which is evident from the Cole-Cole diagram \cite{cole} shown in \ref{fig.8}. The semi-circle in the Cole-Cole
plot characterizes the dielectric data.
 \begin{figure}
\includegraphics[width= 8.25cm]{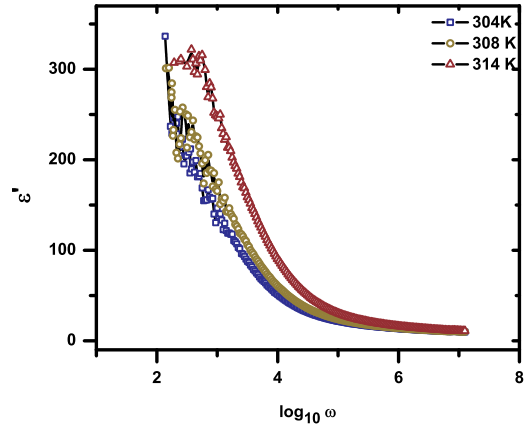}
\caption{(Color online) Variation of real part of the dielectric permittivity ${(\epsilon{'})}$ of the nanocomposite as a function of the frequency at different temperatures.}
\label{fig.6}
   \end{figure}
 \begin{figure}
\includegraphics[width= 8.25cm]{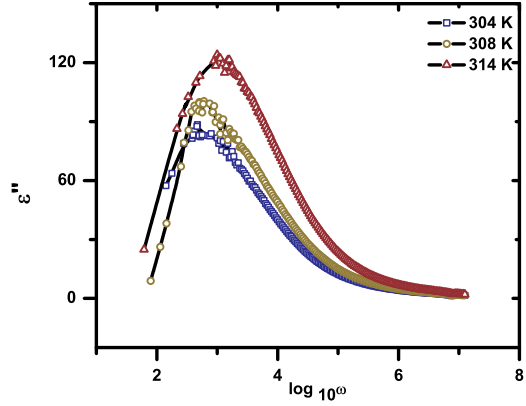}
\caption{(Color online) Variation of imaginary part of the dielectric permittivity ${(\epsilon{''})}$ of the nanocomposite with the frequency at different temperatures.}
\label{fig.7}
   \end{figure}
 \begin{figure}
\includegraphics[width= 8.25cm]{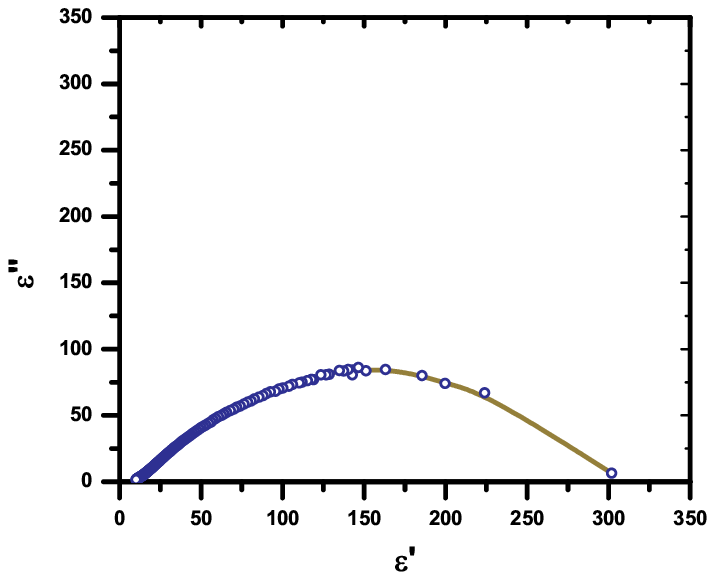}
\caption{(Color online) Cole-Cole diagram of the nanocomposite measured at 308 K.}
\label{fig.8}
   \end{figure}
The dielectric data can be explained on the basis of an inhomogeneous dielectric having laminar structure with different values of dielectric permittivity and electrical 
conductivity. It was shown previously, such systems exhibited dielectric dispersion with a single relaxation time \cite{cole,isard} which is evident from the Cole-Cole diagram.
In the present case, the laminae consist of two-dimensional graphene sheets acting as conducting layer surrounded by the insulating PVA. The analysis of the dielectric data
signifies that the graphene-PVA nanocomposite is electrically inhomogeneous.
\par
\begin{figure}
\includegraphics[width= 8.25cm]{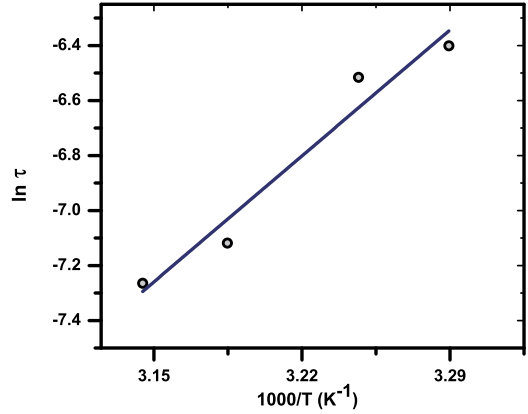}
\caption{(Color online) Arrhenius plot of relaxation times versus 1/T and the respective linear fit for the composite.}
\label{fig.9}
   \end{figure}

From \ref{fig.7} the values of the relaxation times $\tau$ corresponding to the maximum values of ${(\epsilon{''})}$ at different temperatures were 
found out and the results are plotted in \ref{fig.9} as a function of T$^{-1}$. The activation energy of the relaxation process was estimated according to \cite{carbon}
   \begin{equation}
    \textcolor{blue}{\tau=\tau_{o} \exp{(E_{m}/k_{B}T)}}
   \end{equation}
where $E_{m}$ is the activation energy for the relaxation process and T is absolute temperature. From the slope of the straight line the value of 
 the activation energy of 1.29 eV was extracted.

\par
\begin{figure}
\includegraphics[width= 8.25cm]{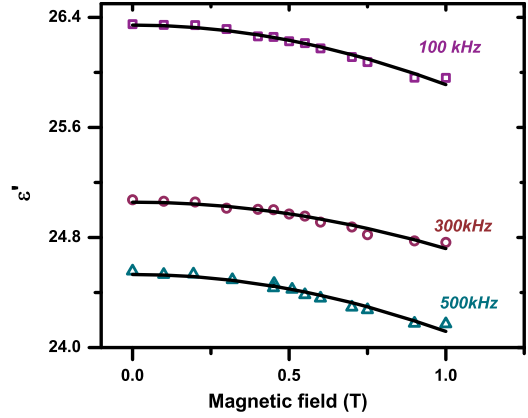}
\caption{(Color online)Variation of the real part of dielectric permittivity ${(\epsilon{'})}$ of the nanocomposite
with magnetic field at different electric field frequencies.}
\label{fig.10}
   \end{figure}

Our system resembles an ideal Maxwell-Wagner dielectric, for which we have investigated a possible magnetodielectric effect. This work has been motivated by a recent 
theoretical work \cite{littlewood}. We had earlier reported this effect in case of nickel nanosheets embedded within the crystal channels of Na-4 mica \cite{selfepl}. 
This is essentially an interface effect, in which the space charges accumulate at the interface of the conducting and non conducting layers. This causes an 
effective dipole moment in the system, reflected in the corresponding dielectric constant of the nanocomposite. By the application of the magnetic field,the charges move 
perpendicular to both the electric and magnetic field directions. The movement of these charges, causes the lowering of the space charge density and the value of 
dielectric constant is decreased. The dielectric permittivity of an inhomogeneous medium consisting of a purely capacitive and a purely resistive region connected in series,
under magnetic field has been given by, \cite{littlewood}
\begin{equation}
 \epsilon_{c}({\omega})={\epsilon}\frac {(1+i{\omega}{\gamma})} {\sqrt{i{\omega}{\gamma}[(1+i{{\omega}{\gamma})}^{2}-{({\omega}{\gamma}{\beta})}^{2}]}}
\end{equation}
where $\epsilon_{c}({\omega})$ is the effective dielectric permittivity, $\omega$ is angular frequency of the applied electric field, $\epsilon$ is the dielectric constant of the
polymer matrix, $\gamma=\epsilon\rho$, $\rho$ is the resistivity of the graphene film, and $\beta=\mu H$; $\mu$ is the carrier mobility and H is the applied magnetic field.
The variation of the real part of the dielectric permittivity of the nanocomposite film with the magnetic field has been shown in \ref{fig.10}.
\begin{figure}
\includegraphics[width= 8.25cm]{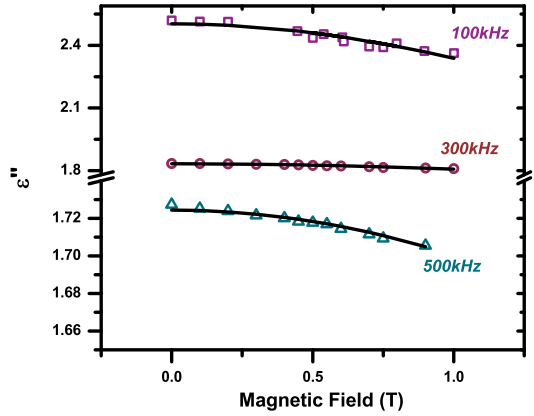}
\caption{(Color online)Variation of the imaginary part of dielectric permittivity ${(\epsilon{''})}$
of the nanocomposite with magnetic field at different electric field frequencies.}
\label{fig.11}
   \end{figure}
As explained above,the dielectric constant decreases with the increase in magnetic field. We have fitted the experimental data with equation (4), for different frequencies
of applied electric field,and the fitting looks satisfactory for all frequencies. In the calculation, we have used $\rho=10^{-6}\Omega-cm$ \cite{geim} for graphene film 
and $\epsilon=2$ for PVA \cite{url}. It is worthy to note that the dielectric constant of the composite decreases with the increasing frequencies,which is a characteristic 
feature of the Maxwell-Wagner polarization effect \cite{hippel}. The variation of the imaginary part of the dielectric constant with the magnetic field has also been shown in 
\ref{fig.11}, and the solid line represents the theoretically fitted curve.
\par
From the magnetic field dependence of the dielectric permittivity of this material one will be tempted to make use of it to probe magnetic field.
 However the effect is admittedly quite small for such an application to be feasible.Use as a magnetic sensor in nanocircuits could still be a possibility.
On the other hand, such materials should be explored for making nano-generators viz;by 
applying an alternating magnetic field a sinusoidal variation of polarization (i.e. bound charge density at electrodes) could be effected. From the latter by a suitable
electrical circuitry a sinusoidal voltage could be generated.Such work will be taken up in our laboratory shortly.  

\section{Conclusion}\label{sec:4}
In summary, graphene sheets were produced by reduction and subsequent exfoliation of graphite oxide powder. PVA solution with dispersed graphene nanosheets were solidified, 
which led to the formation of graphene-PVA nanocomposite films of thickness 120 $\mu m$. Electrical conductivity of the composite was found to be caused by
hopping of carriers between localised states formed at graphne-PVA interface. Dielectric dispersion data showed a Debye like relaxation mechanism.
The nanocomposites exhibited magnetodielectric effect which was explained on the basis of a heterogeneous two-dimensional,two-component composite model.

\section*{Acknowledgement}
  Support for this work was derived from an Indo-Australian Project on nanocomposites for Clean Energy granted by Department of Science and Technology, Govt. of India,
New Delhi. SM thanks University Grants Comission,New Delhi,and DRS thanks Council of Scientific and Industrial Research,New Delhi for awarding Senior Research Fellowships. 
DC thanks Indian National Science Academy, New Delhi, for an Honorary Scientist's position.

\begin{thebibliography}{10}%
\makeatletter
\providecommand \@ifxundefined [1]{%
 \ifx #1\undefined \expandafter \@firstoftwo
 \else \expandafter \@secondoftwo
\fi
}%
\providecommand \@ifnum [1]{%
 \ifnum #1\expandafter \@firstoftwo
 \else \expandafter \@secondoftwo
\fi
}%
\providecommand \enquote [1]{``#1''}%
\providecommand \bibnamefont  [1]{#1}%
\providecommand \bibfnamefont [1]{#1}%
\providecommand \citenamefont [1]{#1}%
\providecommand\href[0]{\@sanitize\@href}%
\providecommand\@href[1]{\endgroup\@@startlink{#1}\endgroup\@@href}%
\providecommand\@@href[1]{#1\@@endlink}%
\providecommand \@sanitize [0]{\begingroup\catcode`\&12\catcode`\#12\relax}%
\@ifxundefined \pdfoutput {\@firstoftwo}{%
 \@ifnum{\z@=\pdfoutput}{\@firstoftwo}{\@secondoftwo}%
}{%
 \providecommand\@@startlink[1]{\leavevmode}%
 \providecommand\@@endlink[0]{}%
}{%
 \providecommand\@@startlink[1]{%
  \leavevmode
  \pdfstartlink
   attr{/Border[0 0 1 ]/H/I/C[0 1 1]}%
   user{/Subtype/Link/A<</Type/Action/S/URI/URI(#1)>>}%
  \relax
 }%
 \providecommand\@@endlink[0]{\pdfendlink}%
}%
\providecommand \url  [0]{\begingroup\@sanitize \@url }%
\providecommand \@url [1]{\endgroup\@href {#1}{\urlprefix}}%
\providecommand \urlprefix [0]{URL }%
\providecommand \Eprint[0]{\href }%
\@ifxundefined \urlstyle {%
  \providecommand \doi [1]{doi:\discretionary{}{}{}#1}%
}{%
  \providecommand \doi [0]{doi:\discretionary{}{}{}\begingroup
  \urlstyle{rm}\Url }%
}%
\providecommand \doibase [0]{http://dx.doi.org/}%
\providecommand \Doi[1]{\href{\doibase#1}}%
\providecommand \bibAnnote [3]{%
  \BibitemShut{#1}%
  \begin{quotation}\noindent
    \textsc{Key:}\ #2\\\textsc{Annotation:}\ #3%
  \end{quotation}%
}%
\providecommand \bibAnnoteFile [2]{%
  \IfFileExists{#2}{\bibAnnote {#1} {#2} {\input{#2}}}{}%
}%
\providecommand \typeout [0]{\immediate \write \m@ne }%
\providecommand \selectlanguage [0]{\@gobble}%
\providecommand \bibinfo [0]{\@secondoftwo}%
\providecommand \bibfield [0]{\@secondoftwo}%
\providecommand \translation [1]{[#1]}%
\providecommand \BibitemOpen[0]{}%
\providecommand \bibitemStop [0]{}%
\providecommand \bibitemNoStop [0]{.\EOS\space}%
\providecommand \EOS [0]{\spacefactor3000\relax}%
\providecommand \BibitemShut [1]{\csname bibitem#1\endcsname}%
\bibitem{geim}%
  \BibitemOpen
  \bibfield{author}{%
  \bibinfo {author} {\bibfnamefont{A.~K.}\ \bibnamefont{Geim}},\ }%
  \bibfield{journal}{%
  \Doi{10.1126/science.1158877}{\bibinfo {journal} {Science}}\ }%
  \textbf{\bibinfo {volume} {324}},\ \bibinfo {pages} {1530} (\bibinfo {year}
  {2009})%
  \bibAnnoteFile{NoStop}{geim}%
\bibitem{novo}%
  \BibitemOpen
  \bibfield{author}{%
  \bibinfo {author} {\bibfnamefont{A.~K.}\ \bibnamefont{Geim}}\ and\ \bibinfo
  {author} {\bibfnamefont{K.~S.}\ \bibnamefont{Novoselov}},\ }%
  \bibfield{journal}{%
  \Doi{10.1038/nmat1849}{\bibinfo {journal} {Nat. Mater.}}\ }%
  \textbf{\bibinfo {volume} {6}},\ \bibinfo {pages} {183} (\bibinfo {year}
  {2007})%
  \bibAnnoteFile{NoStop}{novo}%
\bibitem{stankovich}%
  \BibitemOpen
  \bibfield{author}{%
  \bibinfo {author} {\bibfnamefont{S.}~\bibnamefont{Stankovich}}, \bibinfo
  {author} {\bibfnamefont{D.~A.}\ \bibnamefont{Dikin}}, \bibinfo {author}
  {\bibfnamefont{R.~D.}\ \bibnamefont{Piner}}, \bibinfo {author}
  {\bibfnamefont{K.~A.}\ \bibnamefont{Kohlhaas}}, \bibinfo {author}
  {\bibfnamefont{A.}~\bibnamefont{Kleinhammes}}, \bibinfo {author}
  {\bibfnamefont{Y.}~\bibnamefont{Jia}}, \bibinfo {author}
  {\bibfnamefont{Y.}~\bibnamefont{Wu}}, \bibinfo {author}
  {\bibfnamefont{S.~T.}\ \bibnamefont{Nguyen}},\ and\ \bibinfo {author}
  {\bibfnamefont{R.~S.}\ \bibnamefont{Ruoff}},\ }%
  \bibfield{journal}{%
  \Doi{10.1016/j.carbon.2007.02.034}{\bibinfo {journal} {Carbon}}\ }%
  \textbf{\bibinfo {volume} {45}},\ \bibinfo {pages} {1558} (\bibinfo {year}
  {2007})%
  \bibAnnoteFile{NoStop}{stankovich}%
\bibitem{lotya}%
  \BibitemOpen
  \bibfield{author}{%
  \bibinfo {author} {\bibfnamefont{M.}~\bibnamefont{Lotya}}, \bibinfo {author}
  {\bibfnamefont{Y.}~\bibnamefont{Hernandez}}, \bibinfo {author}
  {\bibfnamefont{P.~J.}\ \bibnamefont{King}}, \bibinfo {author}
  {\bibfnamefont{R.~J.}\ \bibnamefont{Smith}}, \bibinfo {author}
  {\bibfnamefont{V.}~\bibnamefont{Nicolosi}}, \bibinfo {author}
  {\bibfnamefont{L.~S.}\ \bibnamefont{Karlsson}}, \bibinfo {author}
  {\bibfnamefont{F.~M.}\ \bibnamefont{Blighe}}, \bibinfo {author}
  {\bibfnamefont{S.}~\bibnamefont{De}}, \bibinfo {author}
  {\bibfnamefont{Z.}~\bibnamefont{Wang}}, \bibinfo {author}
  {\bibfnamefont{I.~T.}\ \bibnamefont{McGovern}}, ,\ and\ \bibinfo {author}
  {\bibfnamefont{J.~N.~C.}\ \bibnamefont{{Georg S. Duesberg}}},\ }%
  \bibfield{journal}{%
  \Doi{10.1021/ja807449u}{\bibinfo {journal} {J. AM. CHEM. SOC.}}\ }%
  \textbf{\bibinfo {volume} {131}},\ \bibinfo {pages} {3611} (\bibinfo {year}
  {2009})%
  \bibAnnoteFile{NoStop}{lotya}%
\bibitem{cnrjpc}%
  \BibitemOpen
  \bibfield{author}{%
  \bibinfo {author} {\bibfnamefont{K.~S.}\ \bibnamefont{Subrahmanyam}},
  \bibinfo {author} {\bibfnamefont{L.~S.}\ \bibnamefont{Panchakarla}}, \bibinfo
  {author} {\bibfnamefont{A.}~\bibnamefont{Govindaraj}},\ and\ \bibinfo
  {author} {\bibfnamefont{C.~N.~R.}\ \bibnamefont{Rao}},\ }%
  \bibfield{journal}{%
  \Doi{10.1021/jp900791y}{\bibinfo {journal} {J. Phys. Chem. C. Lett.}}\ }%
  \textbf{\bibinfo {volume} {113}},\ \bibinfo {pages} {4257} (\bibinfo {year}
  {2009})%
  \bibAnnoteFile{NoStop}{cnrjpc}%
\bibitem{cnrjmc}%
  \BibitemOpen
  \bibfield{author}{%
  \bibinfo {author} {\bibfnamefont{K.~S.}\ \bibnamefont{Subrahmanyam}},
  \bibinfo {author} {\bibfnamefont{S.~R.~C.}\ \bibnamefont{Vivekchand}},
  \bibinfo {author} {\bibfnamefont{A.}~\bibnamefont{Govindaraj}},\ and\
  \bibinfo {author} {\bibfnamefont{C.~N.~R.}\ \bibnamefont{Rao}},\ }%
  \bibfield{journal}{%
  \Doi{10.1039/b716536f}{\bibinfo {journal} {J. Mater. Chem.}}\ }%
  \textbf{\bibinfo {volume} {18}},\ \bibinfo {pages} {1517} (\bibinfo {year}
  {2008})%
  \bibAnnoteFile{NoStop}{cnrjmc}%
\bibitem{cnrwi}%
  \BibitemOpen
  \bibfield{author}{%
  \bibinfo {author} {\bibfnamefont{C.~N.~R.}\ \bibnamefont{Rao}}, \bibinfo
  {author} {\bibfnamefont{A.}~\bibnamefont{Sood}}, \bibinfo {author}
  {\bibfnamefont{K.~S.}\ \bibnamefont{Subrahmanyam}},\ and\ \bibinfo {author}
  {\bibfnamefont{A.}~\bibnamefont{Govindaraj}},\ }%
  \bibfield{journal}{%
  \Doi{10.1002/anie.200901678}{\bibinfo {journal} {Angew. Chem. Int. Ed.}}\ }%
  \textbf{\bibinfo {volume} {48}},\ \bibinfo {pages} {7752} (\bibinfo {year}
  {2009})%
  \bibAnnoteFile{NoStop}{cnrwi}%
\bibitem{reinanl}%
  \BibitemOpen
  \bibfield{author}{%
  \bibinfo {author} {\bibnamefont{{Alfonso Reina}}}, \bibinfo {author}
  {\bibfnamefont{X.}~\bibnamefont{Jia}}, \bibinfo {author}
  {\bibfnamefont{J.}~\bibnamefont{Ho}}, \bibinfo {author}
  {\bibfnamefont{D.}~\bibnamefont{Nezich}}, \bibinfo {author}
  {\bibfnamefont{H.}~\bibnamefont{Son}}, \bibinfo {author}
  {\bibfnamefont{V.}~\bibnamefont{Bulovic}}, \bibinfo {author}
  {\bibfnamefont{M.~S.}\ \bibnamefont{Dresselhaus}},\ and\ \bibinfo {author}
  {\bibfnamefont{J.}~\bibnamefont{Kong}},\ }%
  \bibfield{journal}{%
  \Doi{10.1021/nl801827v}{\bibinfo {journal} {Nano Lett.}}\ }%
  \textbf{\bibinfo {volume} {9}},\ \bibinfo {pages} {30} (\bibinfo {year}
  {2009})%
  \bibAnnoteFile{NoStop}{reinanl}%
\bibitem{castro}%
  \BibitemOpen
  \bibfield{author}{%
  \bibinfo {author} {\bibfnamefont{A.~H.~C.}\ \bibnamefont{Neto}}, \bibinfo
  {author} {\bibfnamefont{F.}~\bibnamefont{Guinea}}, \bibinfo {author}
  {\bibfnamefont{N.~M.~R.}\ \bibnamefont{Peres}}, \bibinfo {author}
  {\bibfnamefont{K.~S.}\ \bibnamefont{Novoselov}},\ and\ \bibinfo {author}
  {\bibfnamefont{A.~K.}\ \bibnamefont{Geim}},\ }%
  \bibfield{journal}{%
  \Doi{10.1103/RevModPhys.81.109}{\bibinfo {journal} {Rev. Mod. Phys.}}\ }%
  \textbf{\bibinfo {volume} {81}},\ \bibinfo {pages} {109} (\bibinfo {year}
  {2009})%
  \bibAnnoteFile{NoStop}{castro}%
\bibitem{sds}%
  \BibitemOpen
  \bibfield{author}{%
  \bibinfo {author} {\bibfnamefont{S.~D.}\ \bibnamefont{Sarma}}, \bibinfo
  {author} {\bibfnamefont{P.}~\bibnamefont{Kim}}, \bibinfo {author}
  {\bibfnamefont{A.~K.}\ \bibnamefont{Geim}},\ and\ \bibinfo {author}
  {\bibfnamefont{A.~H.}\ \bibnamefont{MacDonald}},\ }%
  \bibfield{journal}{%
  \Doi{10.1016/j.ssc.2007.04.030}{\bibinfo {journal} {Solid State
  Communications}}\ }%
  \textbf{\bibinfo {volume} {143}},\ \bibinfo {pages} {1} (\bibinfo {year}
  {2007})%
  \bibAnnoteFile{NoStop}{sds}%
\bibitem{allor}%
  \BibitemOpen
  \bibfield{author}{%
  \bibinfo {author} {\bibfnamefont{D.}~\bibnamefont{Allor}}, \bibinfo {author}
  {\bibfnamefont{T.~D.}\ \bibnamefont{Cohen}},\ and\ \bibinfo {author}
  {\bibfnamefont{D.~A.}\ \bibnamefont{McGady}},\ }%
  \bibfield{journal}{%
  \Doi{10.1103/PhysRevD.78.096009}{\bibinfo {journal} {Phys. Rev. D}}\ }%
  \textbf{\bibinfo {volume} {78}},\ \bibinfo {pages} {096009} (\bibinfo {year}
  {2008})%
  \bibAnnoteFile{NoStop}{allor}%
\bibitem{shytov}%
  \BibitemOpen
  \bibfield{author}{%
  \bibinfo {author} {\bibfnamefont{A.}~\bibnamefont{Shytov}}, \bibinfo {author}
  {\bibfnamefont{M.~I.}\ \bibnamefont{Katsnelson}},\ and\ \bibinfo {author}
  {\bibfnamefont{L.~S.}\ \bibnamefont{Levitov}},\ }%
  \bibfield{journal}{%
  \Doi{10.1103/PhysRevLett.99.246802}{\bibinfo {journal} {Phys. Rev. Lett.}}\
  }%
  \textbf{\bibinfo {volume} {99}},\ \bibinfo {pages} {246802} (\bibinfo {year}
  {2007})%
  \bibAnnoteFile{NoStop}{shytov}%
\bibitem{young}%
  \BibitemOpen
  \bibfield{author}{%
  \bibinfo {author} {\bibfnamefont{A.~F.}\ \bibnamefont{Young}}\ and\ \bibinfo
  {author} {\bibfnamefont{P.}~\bibnamefont{Kim}},\ }%
  \bibfield{journal}{%
  \Doi{10.1038/NPHYS1198}{\bibinfo {journal} {Nat. Phys.}}\ }%
  \textbf{\bibinfo {volume} {5}},\ \bibinfo {pages} {222} (\bibinfo {year}
  {2009})%
  \bibAnnoteFile{NoStop}{young}%
\bibitem{stander}%
  \BibitemOpen
  \bibfield{author}{%
  \bibinfo {author} {\bibfnamefont{N.}~\bibnamefont{Stander}}, \bibinfo
  {author} {\bibnamefont{{B. Huard}}},\ and\ \bibinfo {author}
  {\bibfnamefont{D.}~\bibnamefont{Goldhaber-Gordon}},\ }%
  \bibfield{journal}{%
  \Doi{10.1103/PhysRevLett.102.026807}{\bibinfo {journal} {Phys. Rev. Lett.}}\
  }%
  \textbf{\bibinfo {volume} {102}},\ \bibinfo {pages} {026807} (\bibinfo {year}
  {2009})%
  \bibAnnoteFile{NoStop}{stander}%
\bibitem{sui}%
  \BibitemOpen
  \bibfield{author}{%
  \bibinfo {author} {\bibfnamefont{Y.}~\bibnamefont{Sui}}, \bibinfo {author}
  {\bibfnamefont{T.}~\bibnamefont{Low}}, \bibinfo {author}
  {\bibfnamefont{M.}~\bibnamefont{Lundstrom}},\ and\ \bibinfo {author}
  {\bibfnamefont{J.}~\bibnamefont{Appenzeller}},\ }%
  \bibfield{journal}{%
  \Doi{10.1021/nl104399z}{\bibinfo {journal} {Nano Lett.}}\ }%
  \textbf{\bibinfo {volume} {11}},\ \bibinfo {pages} {1319} (\bibinfo {year}
  {2011})%
  \bibAnnoteFile{NoStop}{sui}%
\bibitem{watcharotonenl}%
  \BibitemOpen
  \bibfield{author}{%
  \bibinfo {author} {\bibfnamefont{S.}~\bibnamefont{Watcharotone}}, \bibinfo
  {author} {\bibfnamefont{D.~A.}\ \bibnamefont{Dikin}}, \bibinfo {author}
  {\bibfnamefont{S.}~\bibnamefont{Stankovich}}, \bibinfo {author}
  {\bibfnamefont{R.}~\bibnamefont{Piner}}, \bibinfo {author}
  {\bibfnamefont{I.}~\bibnamefont{Jung}}, \bibinfo {author}
  {\bibfnamefont{G.~H.~B.}\ \bibnamefont{Dommett}}, \bibinfo {author}
  {\bibfnamefont{G.}~\bibnamefont{Evmenenko}}, \bibinfo {author}
  {\bibfnamefont{S.-E.}\ \bibnamefont{Wu}}, \bibinfo {author}
  {\bibfnamefont{S.-F.}\ \bibnamefont{Chen}}, \bibinfo {author}
  {\bibfnamefont{C.-P.}\ \bibnamefont{Liu}}, \bibinfo {author}
  {\bibfnamefont{S.~T.}\ \bibnamefont{Nguyen}},\ and\ \bibinfo {author}
  {\bibfnamefont{R.~S.}\ \bibnamefont{Ruoff}},\ }%
  \bibfield{journal}{%
  \Doi{10.1021/nl070477}{\bibinfo {journal} {Nano Lett.}}\ }%
  \textbf{\bibinfo {volume} {7}},\ \bibinfo {pages} {1888} (\bibinfo {year}
  {2007})%
  \bibAnnoteFile{NoStop}{watcharotonenl}%
\bibitem{stankovichnat}%
  \BibitemOpen
  \bibfield{author}{%
  \bibinfo {author} {\bibfnamefont{S.}~\bibnamefont{Stankovich}}, \bibinfo
  {author} {\bibfnamefont{D.~A.}\ \bibnamefont{Dikin}}, \bibinfo {author}
  {\bibfnamefont{G.~H.~B.}\ \bibnamefont{Dommett}}, \bibinfo {author}
  {\bibfnamefont{K.~M.}\ \bibnamefont{Kohlhaas}}, \bibinfo {author}
  {\bibfnamefont{E.~J.}\ \bibnamefont{Zimney}}, \bibinfo {author}
  {\bibfnamefont{E.~A.}\ \bibnamefont{Stach}}, \bibinfo {author}
  {\bibfnamefont{R.~D.}\ \bibnamefont{Piner}}, \bibinfo {author}
  {\bibfnamefont{S.~T.}\ \bibnamefont{Nguyen}},\ and\ \bibinfo {author}
  {\bibfnamefont{R.~S.}\ \bibnamefont{Ruoff}},\ }%
  \bibfield{journal}{%
  \Doi{10.1038/nature04969}{\bibinfo {journal} {Nature}}\ }%
  \textbf{\bibinfo {volume} {442}},\ \bibinfo {pages} {282} (\bibinfo {year}
  {2006})%
  \bibAnnoteFile{NoStop}{stankovichnat}%
\bibitem{xujpcc}%
  \BibitemOpen
  \bibfield{author}{%
  \bibinfo {author} {\bibfnamefont{C.}~\bibnamefont{Xu}}, \bibinfo {author}
  {\bibfnamefont{X.}~\bibnamefont{Wang}},\ and\ \bibinfo {author}
  {\bibfnamefont{J.}~\bibnamefont{Zhu}},\ }%
  \bibfield{journal}{%
  \Doi{10.1021/jp807989b}{\bibinfo {journal} {J. Phys. Chem. C}}\ }%
  \textbf{\bibinfo {volume} {112}},\ \bibinfo {pages} {19841} (\bibinfo {year}
  {2008})%
  \bibAnnoteFile{NoStop}{xujpcc}%
\bibitem{cnrpnas}%
  \BibitemOpen
  \bibfield{author}{%
  \bibinfo {author} {\bibfnamefont{K.~S.}\ \bibnamefont{Subrahmanyam}},
  \bibinfo {author} {\bibfnamefont{P.}~\bibnamefont{Kumar}}, \bibinfo {author}
  {\bibfnamefont{U.}~\bibnamefont{Maitra}}, \bibinfo {author}
  {\bibfnamefont{A.}~\bibnamefont{Govindaraj}}, \bibinfo {author}
  {\bibfnamefont{K.~P. S.~S.}\ \bibnamefont{Hembram}}, \bibinfo {author}
  {\bibfnamefont{U.~V.}\ \bibnamefont{Waghmare}},\ and\ \bibinfo {author}
  {\bibfnamefont{C.~N.~R.}\ \bibnamefont{Rao}},\ }%
  \bibfield{journal}{%
  \Doi{10.1073/pnas.1019542108}{\bibinfo {journal} {Proc. of the Nat. Acad. of
  Sc. of USA}}\ }%
  \textbf{\bibinfo {volume} {108}},\ \bibinfo {pages} {2674} (\bibinfo {year}
  {2011})%
  \bibAnnoteFile{NoStop}{cnrpnas}%
\bibitem{almashatjpcc}%
  \BibitemOpen
  \bibfield{author}{%
  \bibinfo {author} {\bibfnamefont{L.}~\bibnamefont{Al-Mashat}}, \bibinfo
  {author} {\bibfnamefont{K.}~\bibnamefont{Shin}}, \bibinfo {author}
  {\bibfnamefont{K.}~\bibnamefont{Kalantar-zadeh}}, \bibinfo {author}
  {\bibfnamefont{J.~D.}\ \bibnamefont{Plessis}}, \bibinfo {author}
  {\bibfnamefont{S.~H.}\ \bibnamefont{Han}}, \bibinfo {author}
  {\bibfnamefont{R.~W.}\ \bibnamefont{Kojima}}, \bibinfo {author}
  {\bibfnamefont{R.~B.}\ \bibnamefont{Kaner}}, \bibinfo {author}
  {\bibfnamefont{D.}~\bibnamefont{Li}}, \bibinfo {author}
  {\bibfnamefont{X.}~\bibnamefont{Gou}}, \bibinfo {author}
  {\bibfnamefont{S.~J.}\ \bibnamefont{Ippolito}},\ and\ \bibinfo {author}
  {\bibfnamefont{W.}~\bibnamefont{Wlodarski}},\ }%
  \bibfield{journal}{%
  \Doi{10.1021/jp103134u}{\bibinfo {journal} {J. Phys. Chem. C}}\ }%
  \textbf{\bibinfo {volume} {114}},\ \bibinfo {pages} {16168} (\bibinfo {year}
  {2010})%
  \bibAnnoteFile{NoStop}{almashatjpcc}%
\bibitem{hummersjacs}%
  \BibitemOpen
  \bibfield{author}{%
  \bibinfo {author} {\bibfnamefont{W.~S.}\ \bibnamefont{Hummers}}\ and\
  \bibinfo {author} {\bibfnamefont{R.~E.}\ \bibnamefont{Offeman}},\ }%
  \bibfield{journal}{%
  \bibinfo {journal} {J. Am. Chem. Soc.}\ }%
  \textbf{\bibinfo {volume} {80}},\ \bibinfo {pages} {1339} (\bibinfo {year}
  {1958})%
  \bibAnnoteFile{NoStop}{hummersjacs}%
\bibitem{zhoujpcc}%
  \BibitemOpen
  \bibfield{author}{%
  \bibinfo {author} {\bibfnamefont{X.}~\bibnamefont{Zhou}}, \bibinfo {author}
  {\bibfnamefont{X.}~\bibnamefont{Huang}}, \bibinfo {author}
  {\bibfnamefont{X.}~\bibnamefont{Qi}}, \bibinfo {author}
  {\bibfnamefont{S.}~\bibnamefont{Wu}}, \bibinfo {author}
  {\bibfnamefont{C.}~\bibnamefont{Xue}}, \bibinfo {author}
  {\bibfnamefont{F.~Y.~C.}\ \bibnamefont{Boey}}, \bibinfo {author}
  {\bibfnamefont{Q.}~\bibnamefont{Yan}}, \bibinfo {author}
  {\bibfnamefont{P.}~\bibnamefont{Chen}},\ and\ \bibinfo {author}
  {\bibfnamefont{H.}~\bibnamefont{Zhang}},\ }%
  \bibfield{journal}{%
  \Doi{10.1021/jp903821n}{\bibinfo {journal} {J. Phys. Chem. C}}\ }%
  \textbf{\bibinfo {volume} {113}},\ \bibinfo {pages} {10842} (\bibinfo {year}
  {2009})%
  \bibAnnoteFile{NoStop}{zhoujpcc}%
\bibitem{kovtycm}%
  \BibitemOpen
  \bibfield{author}{%
  \bibinfo {author} {\bibfnamefont{N.~I.}\ \bibnamefont{Kovtyukhova}}, \bibinfo
  {author} {\bibfnamefont{P.~J.}\ \bibnamefont{Ollivier}}, \bibinfo {author}
  {\bibfnamefont{B.~R.}\ \bibnamefont{Martin}}, \bibinfo {author}
  {\bibfnamefont{T.~E.}\ \bibnamefont{Mallouk}}, \bibinfo {author}
  {\bibfnamefont{S.~A.}\ \bibnamefont{Chizhik}}, \bibinfo {author}
  {\bibfnamefont{E.~V.}\ \bibnamefont{Buzaneva}},\ and\ \bibinfo {author}
  {\bibfnamefont{A.~D.}\ \bibnamefont{Gorchinskiy}},\ }%
  \bibfield{journal}{%
  \Doi{10.1021/cm981085u}{\bibinfo {journal} {Chem. Mater.}}\ }%
  \textbf{\bibinfo {volume} {11}},\ \bibinfo {pages} {771} (\bibinfo {year}
  {1999})%
  \bibAnnoteFile{NoStop}{kovtycm}%
\bibitem{dannn}%
  \BibitemOpen
  \bibfield{author}{%
  \bibinfo {author} {\bibfnamefont{D.}~\bibnamefont{Li}}, \bibinfo {author}
  {\bibfnamefont{M.~B.}\ \bibnamefont{M{\"u}ller}}, \bibinfo {author}
  {\bibfnamefont{S.}~\bibnamefont{Gilje}}, \bibinfo {author}
  {\bibfnamefont{R.~B.}\ \bibnamefont{Kaner}},\ and\ \bibinfo {author}
  {\bibfnamefont{G.~G.}\ \bibnamefont{Wallace}},\ }%
  \bibfield{journal}{%
  \Doi{10.1038/nnano.2007.451}{\bibinfo {journal} {Nat. Nanotech.}}\ }%
  \textbf{\bibinfo {volume} {3}},\ \bibinfo {pages} {101} (\bibinfo {year}
  {2008})%
  \bibAnnoteFile{NoStop}{dannn}%
\bibitem{mott}%
  \BibitemOpen
  \bibfield{author}{%
  \bibinfo {author} {\bibfnamefont{N.~F.}\ \bibnamefont{Mott}}\ and\ \bibinfo
  {author} {\bibfnamefont{E.}~\bibnamefont{Davis}},\ }%
  \emph{\bibinfo {title} {Electronic processes in non-crystalline materials}}\
  (\bibinfo {publisher} {Clarendon Press},\ \bibinfo {address} {Oxford, UK},\
  \bibinfo {year} {1971})%
  \bibAnnoteFile{NoStop}{mott}%
\bibitem{hippel}%
  \BibitemOpen
  \bibfield{author}{%
  \bibinfo {author} {\bibfnamefont{A.~R.}\ \bibnamefont{von Hippel}},\ }%
  \emph{\bibinfo {title} {Dielectrics and Waves}}\ (\bibinfo {publisher} {The
  MIT Press},\ \bibinfo {address} {Massechusets,USA},\ \bibinfo {year} {1966})%
  \bibAnnoteFile{NoStop}{hippel}%
\bibitem{cole}%
  \BibitemOpen
  \bibfield{author}{%
  \bibinfo {author} {\bibfnamefont{K.}~\bibnamefont{Cole}}\ and\ \bibinfo
  {author} {\bibfnamefont{R.}~\bibnamefont{Cole}},\ }%
  \bibfield{journal}{%
  \Doi{10.1063/1.1750906}{\bibinfo {journal} {J.Chem.Phys.}}\ }%
  \textbf{\bibinfo {volume} {9}},\ \bibinfo {pages} {341} (\bibinfo {year}
  {1941})%
  \bibAnnoteFile{NoStop}{cole}%
\bibitem{isard}%
  \BibitemOpen
  \bibfield{author}{%
  \bibinfo {author} {\bibfnamefont{J.}~\bibnamefont{Isard}},\ }%
  \bibfield{journal}{%
  \bibinfo {journal} {Proc.Inst.Electr.Eng.}\ }%
  \textbf{\bibinfo {volume} {109B}},\ \bibinfo {pages} {440} (\bibinfo {year}
  {1962})%
  \bibAnnoteFile{NoStop}{isard}%
\bibitem{carbon}%
  \BibitemOpen
  \bibfield{author}{%
  \bibinfo {author} {\bibfnamefont{Y.}~\bibnamefont{Xi}}, \bibinfo {author}
  {\bibfnamefont{Y.}~\bibnamefont{Bin}}, \bibinfo {author}
  {\bibfnamefont{C.}~\bibnamefont{Chiang}},\ and\ \bibinfo {author}
  {\bibfnamefont{M.}~\bibnamefont{Matsuo}},\ }%
  \bibfield{journal}{%
  \Doi{10.1016/j.carbon.2007.01.019}{\bibinfo {journal} {Carbon}}\ }%
  \textbf{\bibinfo {volume} {45}},\ \bibinfo {pages} {1302} (\bibinfo {year}
  {2007})%
  \bibAnnoteFile{NoStop}{carbon}%
\bibitem{littlewood}%
  \BibitemOpen
  \bibfield{author}{%
  \bibinfo {author} {\bibfnamefont{M.~M.}\ \bibnamefont{Parish}}\ and\ \bibinfo
  {author} {\bibfnamefont{P.~B.}\ \bibnamefont{Littlewood}},\ }%
  \bibfield{journal}{%
  \Doi{10.1103/PhysRevLett.101.166602}{\bibinfo {journal} {Phys. Rev. Lett.}}\
  }%
  \textbf{\bibinfo {volume} {101}},\ \bibinfo {pages} {166602} (\bibinfo {year}
  {2008})%
  \bibAnnoteFile{NoStop}{littlewood}%
\bibitem{selfepl}%
  \BibitemOpen
  \bibfield{author}{%
  \bibinfo {author} {\bibfnamefont{S.}~\bibnamefont{Mitra}}, \bibinfo {author}
  {\bibfnamefont{A.}~\bibnamefont{Mandal}}, \bibinfo {author}
  {\bibfnamefont{A.}~\bibnamefont{Datta}}, \bibinfo {author}
  {\bibfnamefont{S.}~\bibnamefont{Banerjee}},\ and\ \bibinfo {author}
  {\bibfnamefont{D.}~\bibnamefont{Chakravorty}},\ }%
  \bibfield{journal}{%
  \Doi{10.1209/0295-5075/92/26003}{\bibinfo {journal} {Euro. Phys. Lett.}}\ }%
  \textbf{\bibinfo {volume} {92}},\ \bibinfo {pages} {26003} (\bibinfo {year}
  {2010})%
  \bibAnnoteFile{NoStop}{selfepl}%
\bibitem{url}%
  \BibitemOpen
  \bibfield{journal}{%
  \bibinfo {journal}
  {https://www.clippercontrols.com/pages/dielectric-constant-values}}%
   (\bibinfo {year} {accessed March 11,2011})%
  \bibAnnoteFile{NoStop}{url}%
\end{thebibliography}
%
 

\end{document}